\documentstyle[eqsecnum,aps,epsfig]{revtex}
\begin{document}
\draft
\title{Nearly horizon skimming orbits of Kerr black holes}
\author{Scott A.\ Hughes}
\address{
Institute for Theoretical Physics, University of California,
Santa Barbara, CA 93103\\
Theoretical Astrophysics, California Institute of Technology,
Pasadena, CA 91125\\
}
\maketitle
\begin{abstract}
An unusual set of orbits about extreme Kerr black holes resides at the
Boyer-Lindquist radius $r = M$, the coordinate of the hole's event
horizon.  These ``horizon skimming'' orbits have the property that
their angular momentum $L_z$ {\it increases} with inclination angle,
opposite to the familiar behavior one encounters at larger radius.  In
this paper, I show that this behavior is characteristic of a larger
family of orbits, the ``nearly horizon skimming'' (NHS) orbits.  NHS
orbits exist in the very strong field of any black hole with spin
$a\agt 0.952412M$.  Their unusual behavior is due to the locking of
particle motion near the event horizon to the hole's spin, and is
therefore a signature of the Kerr metric's extreme strong field.  An
observational hallmark of NHS orbits is that a small body spiraling
into a Kerr black hole due to gravitational-wave emission will be
driven into orbits of progressively smaller inclination angle, toward
the equator.  This is in contrast to the ``normal'' behavior.  For
circular orbits, the change in inclination is very small, and unlikely
to be of observational importance.  I argue that the change in
inclination may be considerably larger when one considers the
evolution of inclined eccentric orbits.  If this proves correct, then
the gravitational waves produced by evolution through the NHS regime
may constitute a very interesting and important probe of the
strong-field nature of rotating black holes.
\end{abstract}

\section{Introduction}
\label{sec:intro}

The space-based gravitational-wave detector LISA {\cite{LISA}} is
being designed to make very precise measurements of the
characteristics of black hole spacetimes.  One source that is
particularly well-suited for such measurements is the inspiral of a
small (mass $\mu = 1 - 10\,M_\odot$) compact body into a massive ($M =
10^{5-7}\,M_\odot$) Kerr black hole.  Depending upon the values of
$M$, $\mu/M$, and the hole's spin $a$, such an inspiral will spend
several months to several years in LISA's frequency band
{\cite{finn_thorne}}, radiating $10^5 - 10^6$ gravitational-wave
cycles.  By accurately measuring these cycles, LISA should be able to
build a ``map'' {\cite{fintan_measure,hughes_map}} of the spacetime,
testing in detail the predictions of general relativity.

As the community begins developing strategies for analyzing LISA's
datastream, it is important to revisit and carefully analyze the
sources one expects to measure.  For extreme mass ratio inspirals,
this means understanding more deeply the character of Kerr black hole
orbits and the nature of gravitational radiation reaction deep in the
Kerr metric's strong field.  Of particular interest are features that
might simplify data analysis (which is likely to be very difficult,
given the many cycles that must be tracked) or that might constitute a
strong signature of the spacetime.  In this paper, I analyze what
might be such a feature --- a unique signature of the inspiral of a
body through the extreme strong field of rapidly rotating black holes.

The key piece of this analysis was first discussed by Wilkins.
Reference {\cite{wilkins}} contains a detailed examination of circular
orbits of extreme ($a = M$)\footnote{I use units where $G = 1 = c$.}
Kerr black holes.  (In this context, ``circular orbit'' means ``orbit
of constant Boyer-Lindquist coordinate radius''.)  One very
interesting result given in {\cite{wilkins}} is the existence of
``horizon skimming'' orbits.  These are circular orbits of varying
inclination angle at the same coordinate radius as the event horizon,
$r = M$.  (Despite being at the same radial coordinate, one can show
that these orbits have distinct proper separation, and in particular
lie outside the event horizon; see Ref.\ {\cite{bpt}}, particularly
Fig.\ 2.)  An extremely interesting feature of the horizon skimming
orbits is that as an orbit's inclination angle $\iota$ is increased,
its angular momentum component $L_z$ likewise increases: $\partial
L_z/\partial\iota > 0$ for the horizon skimming orbits.  This property
holds over a sizable range of radius, out to $r \simeq 1.8 M$.  I will
call the full set of orbits for which $\partial L_z/\partial\iota > 0$
the ``nearly horizon skimming'' (NHS) orbits.  This defining property
of NHS orbits is opposite to weak-field intuition.  For example, in
Newtonian theory, orbits at constant radius have $L_z = |{\vec
L}|\cos\iota$ (where $|\vec L|$ is the same for all orbits at radius
$r$), which decreases as $\iota$ increases.

Intuition from Newtonian theory is highly suspect in the strong field
of black holes.  However, the true behavior of $L_z$ as a function of
$\iota$ is qualitatively the same as in Newtonian theory over a wide
range of orbital radii and spins.  For example, when $a = 0.95 M$,
$\partial L_z/\partial\iota < 0$ for all circular orbits, so there are
no NHS orbits when $a = 0.95 M$.  It turns out that NHS orbits can
only exist when $a\agt 0.952412M$.  This is the smallest spin for
which stable orbits come close enough to the hole's event horizon that
$\partial L_z/\partial\iota$ can switch sign: the property $\partial
L_z/\partial\iota > 0$ arises because, very close to the black hole,
all physical processes become ``locked'' or ``frozen'' to the hole's
event horizon {\cite{membrane}}.  In particular, their orbital motion
locks to the horizon's spin.  This locking dominates the ``Keplerian''
tendency of an orbit to move more quickly at smaller radius ($v_{\rm
Kepler} = \sqrt{M/r}$), forcing a body to actually slow in the
innermost orbits.  The locking is particularly strong for the
most-bound orbits.  As I show below, the least-bound orbits do not
strongly lock to the black hole's spin until they have very nearly
reached the innermost orbit.  The NHS orbit's $\partial
L_z/\partial\iota$ behavior follows from the fact that the most-bound
orbit locks to the horizon more readily than the least-bound orbit.

This behavior could have interesting observational consequences.  It
has been well-understood for some time that the inclination angle of
an inspiraling body increases due to gravitational-wave emission
{\cite{fintan_circ,paperI}}.  Gravitational waves carry $L_z$ away
from the orbit, so that $dL_z/dt < 0$.  Since ``normal'' orbits have
$\partial L_z/\partial\iota < 0$, it follows that $d\iota/dt > 0$.  If
$\partial L_z/\partial\iota$ switches sign, then $d\iota/dt$ will
switch sign as well: an inspiraling body will evolve toward an
equatorial orbit.  If the change in $\iota$ is large, it could have a
large effect on the gravitational waveform.  For example, the
spin-orbit modulation of the wave's amplitude and phase is due to
motion in the $\theta$ coordinate.  This modulation will be reduced as
the body's $\theta$ motion reduces.

Since the size of NHS orbits is significant near the ``astrophysically
maximal'' value $a = 0.998 M$ (the value at which a hole's spin tends
to be buffered due to photon capture from thin disk accretion; see
Ref.\ {\cite{thorne_spin}}), astrophysical black holes might spin
quickly enough for NHS orbits to play some important role.  This
motivates a careful analysis to see what role, if any, NHS orbits
might play in gravitational-wave sources.  I use the code described in
Ref.\ {\cite{paperI}} (which uses the Teukolsky and Sasaki-Nakamura
equations {\cite{teukolsky,sasaknak}} to compute the flux of energy
and $L_z$ carried away from the orbit by gravitational waves) to study
how a small body's motion evolves as it spirals through the NHS
region.  By computing the change $(dr/dt,d\iota/dt)$ at a large number
of points, it is straightforward to construct the inspiral trajectory
for a small body {\cite{paperII}}.  I find that the total change in
inclination angle as a body spirals through the NHS region is very
small --- at most, $\delta\iota\simeq 1^\circ - 2^\circ$.  (See Fig.\
{\ref{fig:nhs_seq}}; note that the {\it shape} of the curves in this
figure are independent of the values of $\mu$ and $M$, although the
timescales strongly depend on $\mu$ and $M$.)

The inspiral code relies on the fact that circular orbits remain
circular as they evolve on an adiabatic timescale
{\cite{mino_circ,fintan_gen,dan_amos}}.  It is thus explicitly
restricted to the evolution of circular orbits, and cannot say
anything about the evolution of eccentric orbits, which are much more
realistic as LISA sources {\cite{sigurdsson_rees,sigurdsson}}.  Based
on the leading order corrections to $d\iota/dt$ seen from a
post-Newtonian analysis {\cite{fintan_gen}}, I speculate that the
change in $\iota$ might be much larger when eccentric, inclined orbits
evolve through the strong field.  Verifying this, however, will
require a strong-field radiation reaction formalism that can evolve
generic Kerr orbits: orbits that are inclined {\it and} eccentric.
Such a formalism may need to be based on a local radiation reaction
force {\cite{rad_react_cabal}}, although recent discussion suggests it
may be possible to evolve generic orbits using gravitational-wave
fluxes alone {\cite{wolfgang_eanna}}.  The possibility that inspiral
through the extreme strong field may leave an observationally
significant imprint on the system's gravitational waveform will
hopefully motivate future activity and progress on this problem.

Throughout this paper, the quantities $t$, $r$, $\theta$, and $\phi$
refer to the Boyer-Lindquist coordinates.  A prime on any quantity
denotes $\partial/\partial r$.  (Note that in Ref.\ {\cite{paperI}} I
erroneously wrote that prime denotes $d/dr$.)  Section
{\ref{sec:circular}} reviews the properties of circular Kerr orbits,
providing formulas that are useful for describing their conserved
quantities $E$, $L_z$, and $Q$ in the very strong field.  These are
used in Sec.\ {\ref{sec:nhs}} to study the NHS orbits.  The NHS orbits
are developed and mapped out as functions of spin and radius in Sec.\
{\ref{subsec:nhs_overview}}.  Section {\ref{subsec:nhs_zamo}} then
re-examines them from the viewpoint of the ``zero angular momentum
observer'', or ZAMO.  The ZAMO makes local measurements of the orbital
properties, and can see that their defining behavior $\partial
L_z/\partial\iota > 0$ arises due to the locking of NHS orbits to the
black hole's spin.  In Sec.\ {\ref{sec:radiate}}, I examine the
trajectory of a body that is inspiraling through the NHS region under
gravitational-wave emission.  Here I show the very small change in
$\iota$ as the body spirals in, and argue that eccentricity might
impact this result greatly.  Some concluding discussion is given in
Sec.\ {\ref{sec:conclusion}}.

\section{Circular orbits of Kerr black holes}
\label{sec:circular}

Geodesic orbits of a Kerr black hole with mass $M$ and spin per unit
mass $a$ are governed by the following four equations {\cite{mtw}}:
\begin{mathletters}
\begin{eqnarray}
\Sigma^2\left({dr\over d\tau}\right)^2 && =\left[E(r^2+a^2)
- a L_z\right]^2- \Delta\left[r^2 + (L_z - a E)^2 +
Q\right] \equiv R\;,\label{eq:rdot}\\
\Sigma^2\left({d\theta\over d\tau}\right)^2 && = Q - \cot^2\theta L_z^2
-a^2\cos^2\theta(1 - E^2)\equiv\Theta^2\;,\label{eq:thetadot}\\
\Sigma\left({d\phi\over d\tau}\right) && =
\csc^2\theta L_z + aE\left({r^2+a^2\over\Delta} - 1\right) -
{a^2L_z\over\Delta}\;,\label{eq:phidot}\\
\Sigma\left({dt\over d\tau}\right) && =
E\left[{(r^2+a^2)^2\over\Delta} - a^2\sin^2\theta\right] +
aL_z\left(1 - {r^2+a^2\over\Delta}\right)\;.\label{eq:tdot}
\label{eq:geodesiceqns}
\end{eqnarray}
\end{mathletters}
The quantities $E$, $L_z$, and $Q$ (``energy'', ``$z$-component of
angular momentum'', and ``Carter constant'') specify a family of
orbits, and are conserved along any orbit of the family.  Here,
$\Sigma = r^2 + a^2\cos^2\theta$ and $\Delta = r^2 - 2 M r + a^2$.
Eqs.\ (\ref{eq:rdot}) and (\ref{eq:thetadot}) have been divided by
$\mu^2$, and Eqs.\ (\ref{eq:phidot}) and (\ref{eq:tdot}) by $\mu$
(where $\mu$ is the mass of a small body in an orbit); $E$, $L_z$, and
$Q$ are thus the specific energy, angular momentum and Carter
constant.  Also, $a \ge 0$; prograde and retrograde orbits are
distinguished by the orbit's tilt angle rather than the sign of the
hole's spin.

A circular orbit must satisfy $R = 0$, $R' = 0$; to be stable, it also
must satisfy $R'' < 0$.  These conditions are met for some set of
orbits everywhere outside the innermost stable circular orbit, or
ISCO.  The ISCO lies at {\cite{bpt}}
\begin{eqnarray}
r_{\rm ISCO}/M &=& 3 + Z_2 - \left[(3 - Z_1)(3 + Z_1 + 2
Z_2)\right]^{1/2}\;,\\
\label{eq:risco}
Z_1 &=& 1 + \left[1 - (a/M)^2\right]^{1/3}\left[(1 + a/M)^{1/3} +
(1 - a/M)^{1/3}\right]\;,\\
\label{eq:Z1}
Z_2 &=& \left[3 (a/M)^2 + Z_1^2\right]^{1/2}\;.
\label{eq:Z2}
\end{eqnarray}
The ISCO varies from $r = 6M$ for a Schwarzschild black hole to $r =
M$ for an extreme Kerr hole.

At all $r \ge r_{\rm ISCO}$ there exists a family of circular orbits,
each member having a different inclination angle $\iota$.  We are
interested in parameterizing these orbits as functions of $r$ and
$\iota$.  Consider first the weak-field limit, $r \gg r_{\rm ISCO}$.
Ryan {\cite{fintan_circ}} has provided formulas which, with some
manipulation, give $E$, $L_z$, and $Q$ as functions of $r$ and
$\iota$:
\begin{eqnarray}
E &=& 1 - {M\over2r} - 2{a\over M}\left({M\over
r}\right)^{3/2}\cos\iota\;,
\label{eq:E_weak}\\
\left(L_z^2 + Q\right)^{1/2} &=& \sqrt{r M}\left[1 - 3{a\over M}
\left({M\over r}\right)^{3/2}\cos\iota\right]\;.
\label{eq:lzsqr_Q_weak}
\end{eqnarray}
[Note there is a sign error in Eq.\ (7) of Ref.\ {\cite{fintan_circ}},
as can be seen by taking the zero eccentricity limit of Eq.\ (6) of
Ref.\ {\cite{fintan_gen}}.]  One can then get $L_z$ from the
definition of the inclination angle:
\begin{equation}
\cos\iota = {L_z\over\sqrt{L_z^2 + Q}}\;.
\label{eq:cosiota}
\end{equation}
Two features of these formulas are particularly noteworthy.  First,
energy monotonically increases as $\iota$ increases: $\partial
E/\partial\iota > 0$ for all parameters.  This turns out to be true
everywhere, not just in the weak field.  Second, $L_z$ monotically
decreases as $\iota$ increases provided we don't abuse the
applicability of Eqs.\ (\ref{eq:lzsqr_Q_weak}) and (\ref{eq:cosiota}):
$\partial L_z/\partial\iota < 0$ except when $r/M \alt \left(6
a/M\right)^{2/3}$.  This range is not even close to the weak-field, so
there is no reason to believe that this result is at all physically
relevant.  Nonetheless, it foreshadows the behavior of the nearly
horizon skimming orbits.

Turn now from the weak field to the strong field.  As is conventional
{\cite{fintan_circ,paperI,fintan_gen}}, I will use Eq.\
(\ref{eq:cosiota}) to define the inclination angle\footnote{As
discussed in Ref.\ {\cite{paperI}}, this angle doesn't necessarily
accord with intuitive notions of inclination angle.  For example,
except when $a = 0$, $\iota$ is {\it not} the angle at which most
observers would see the small body cross the equatorial plane.} even
in this regime.  The {\it most-bound} orbit\footnote{In this paper,
the terms ``most-bound'', ``least-bound'' and ``marginally-bound''
describe orbits at each radius.  This contrasts with the usage in,
{\it e.g.}, Ref.\ {\cite{bpt}} where these terms refer to properties
of {\it all} orbits, regardless of radius.} (the orbit with the
smallest orbital energy) at each radius is the prograde, equatorial
orbit.  Its constants are given by {\cite{bpt}}
\begin{eqnarray}
E^{\rm mb} &=& {{1 - 2 v^2 + q v^3}\over\sqrt{1 - 3 v^2 + 2 q
v^3}}\;,\\
\label{eq:Emb}
L_z^{\rm mb} &=& r v{{1 - 2 q v^3 + q^2 v^4}\over\sqrt{1 - 3 v^2 + 2 q
v^3}}\;,\\
\label{eq:Lzmb}
Q^{\rm mb} &=& 0\;,
\label{eq:Qmb}
\end{eqnarray}
where $v \equiv \sqrt{M/r}$ and $q \equiv a/M$.  At fixed radius, the
orbital energy increases as the tilt increases from the most-bound
orbit at $\iota = 0^\circ$ to the {\it least-bound} orbit.  The
least-bound orbit's characteristics depend upon $r$ and the black
hole's spin.  If $r \ge r_{\rm ret}$, where {\cite{bpt}}
\begin{equation}
r_{\rm ret}/M = 3 + Z_2 + \left[(3 - Z_1)(3 + Z_1 +
2 Z_2)\right]^{1/2}\;,
\label{eq:rret}
\end{equation}
then the least-bound orbit is just the retrograde, equatorial orbit.
This orbit has $Q = 0$ and $\iota = 180^\circ$; expressions for its
energy and angular momentum can be found in Ref.\ {\cite{bpt}}.  For
radii $r_{\rm ISCO} \le r \le r_{\rm ret}$, the least-bound orbit is
the {\it marginally-stable} orbit: the orbit which satisfies $R = 0$,
$R' = 0$, and $R'' = 0$.  This orbit has the maximum allowed
inclination angle $\iota_{\rm max}$ at that radius.  Any orbit tilted
at a larger angle is unstable to small perturbations and will quickly
plunge into the black hole.

For the rest of this paper, I will focus on the extreme strong field
of rapidly rotating black holes.  The orbits of interest are well
inside the radius of the retrograde orbit.  Hence, the energy $E^{\rm
lb}$, angular momentum $L_z^{\rm lb}$, and Carter constant $Q^{\rm
lb}$ of the least-bound orbit will be determined by numerically
solving the equations $R = 0$, $R' = 0$, $R'' = 0$.

To compute the properties of a circular orbit, pick two of its
constants --- {\it e.g.}, the orbit's radius $r$ and angular momentum
$L_z$ --- and solve $R = 0 = R'$ to find the other two.  This yields
the following solution for $E(r,L_z)$ and $Q(r,L_z)$ {\cite{paperI}}:
\begin{eqnarray}
E(r,L_z) &&= {{a^2L_z^2(r-M) + r\Delta^2}
\over{a L_z M\left(r^2 - a^2\right) \pm \Delta\sqrt{r^5(r-3M)
+ a^4r(r+M)+a^2r^2(L_z^2-2Mr+2r^2)]}}},
\label{eq:EofLzr}\\
Q(r,L_z) &&= {\left[(a^2 + r^2) E(r,L_z) - a L_z\right]^2\over \Delta}
- \left[r^2 + a^2 E(r,L_z)^2 - 2 a E(r,L_z) L_z + L_z^2\right]\;.
\label{eq:QofLzr}
\end{eqnarray}
There is a sign choice in the denominator of Eq.\ (\ref{eq:EofLzr}).
In general, only one choice is physically meaningful at a given value
of $r$.  The argument of the square root in the denominator of Eq.\
(\ref{eq:EofLzr}) goes to zero at some radius $r_{\rm branch}(a)$; the
plus sign corresponds to $r \ge r_{\rm branch}(a)$, and the minus sign
to $r \le r_{\rm branch}(a)$.  In Ref.\ {\cite{paperI}}, Eq.\
(\ref{eq:EofLzr}) was used with the plus sign only since the focus in
that paper was on comparatively large radius [in all cases, $r_{\rm
branch}(a)$ is close to $2M$].  In this work, since I will focus on
the extreme strong field, {\it both} signs are needed.

Rather than using Eq.\ (\ref{eq:EofLzr}), I will avoid this sign
ambiguity by starting with $r$ and $E$, and re-solving the system $R =
0 = R'$ for $L_z$ and $Q$.  Then,
\begin{eqnarray}
L_z(r,E) &&= {E M (r^2 - a^2) - \Delta\sqrt{r^2(E^2 - 1) + r M}\over
{a(r - M)}}\;,
\label{eq:LzofEr}\\
Q(r,E) &&= {\left[(a^2 + r^2)E - a L_z(r,E)\right]^2\over \Delta}
- \left[r^2 + a^2 E - 2 a E L_z(r,E) + L_z(r,E)^2\right]\;.
\label{eq:QofEr}
\end{eqnarray}
(There exists a second solution for $L_z$ which has a $+$ sign in
front of the $\Delta$, but it is not physically meaningful.)  Note
that Eq.\ (\ref{eq:LzofEr}) does not behave well as $a \to 0$.  This
is because of a degeneracy in this limit: knowledge of any three of
the parameters $E$, $r$, $L_z$, and $Q$ suffices to determine the
orbit (because of spherical symmetry).  Since this paper deals with
$a\sim M$, this issue is irrelevant here.

Assembling strong-field orbits now reduces to a simple recipe.  First,
pick an orbital radius.  Allow the orbital energy to vary from $E^{\rm
mb}$ [Eq.\ (\ref{eq:Emb})] to $E^{\rm lb}$ (found by solving the
system $R = 0$, $R' = 0$, $R'' = 0$).  For each energy, find $L_z$ and
$Q$ with Eqs.\ (\ref{eq:LzofEr}) and (\ref{eq:QofEr}).  Parameterize
each orbit by its inclination angle $\iota$ [Eq.\ (\ref{eq:cosiota})].
Repeat at a new radius.

\section{Nearly horizon-skimming orbits}
\label{sec:nhs}

\subsection{Overview}
\label{subsec:nhs_overview}

Consider for a moment the extreme Kerr limit, $a = M$.  From Eq.\
(\ref{eq:risco}), the ISCO is located at $r = M$, which is also the
coordinate of the event horizon.  It is not difficult to show that
there exists a set of orbits at this radius, with the parameters
\begin{eqnarray}
2M/\sqrt{3} &\le& L_z \le \sqrt{2}M\;,\\
E &=& L_z/2M\;,\\
Q &=& 3L_z^2/4 - M^2\;.
\label{eq:horiz_skimming}
\end{eqnarray}
These are the horizon skimming orbits.  They obey $\partial
L_z/\partial\iota > 0$, similar to the behavior seen when the
weak-field results for $L_z$ are pushed into the very strong field.
This is opposite to the typical behavior, as exemplified by the
correct usage of Eqs.\ (\ref{eq:lzsqr_Q_weak}) and (\ref{eq:cosiota}).

Focus for now on the most-bound and least-bound orbits.  These orbits
bound the behavior of all orbits at each radius.  As described in
Sec.\ {\ref{sec:circular}}, it is simple to solve for $L_z^{\rm lb}$
as a function of radius, at least numerically.  Doing so for $a = M$,
we find that there is a region stretching to $r \simeq 1.8 M$ in which
$L_z^{\rm lb} \ge L_z^{\rm mb}$.  The orbits in this domain have the
same dynamical characteristics as Wilkins' horizon skimming orbits, so
we shall call them ``nearly horizon skimming'' (NHS) orbits.

\begin{figure}[ht]
\begin{center}
\epsfig{file = 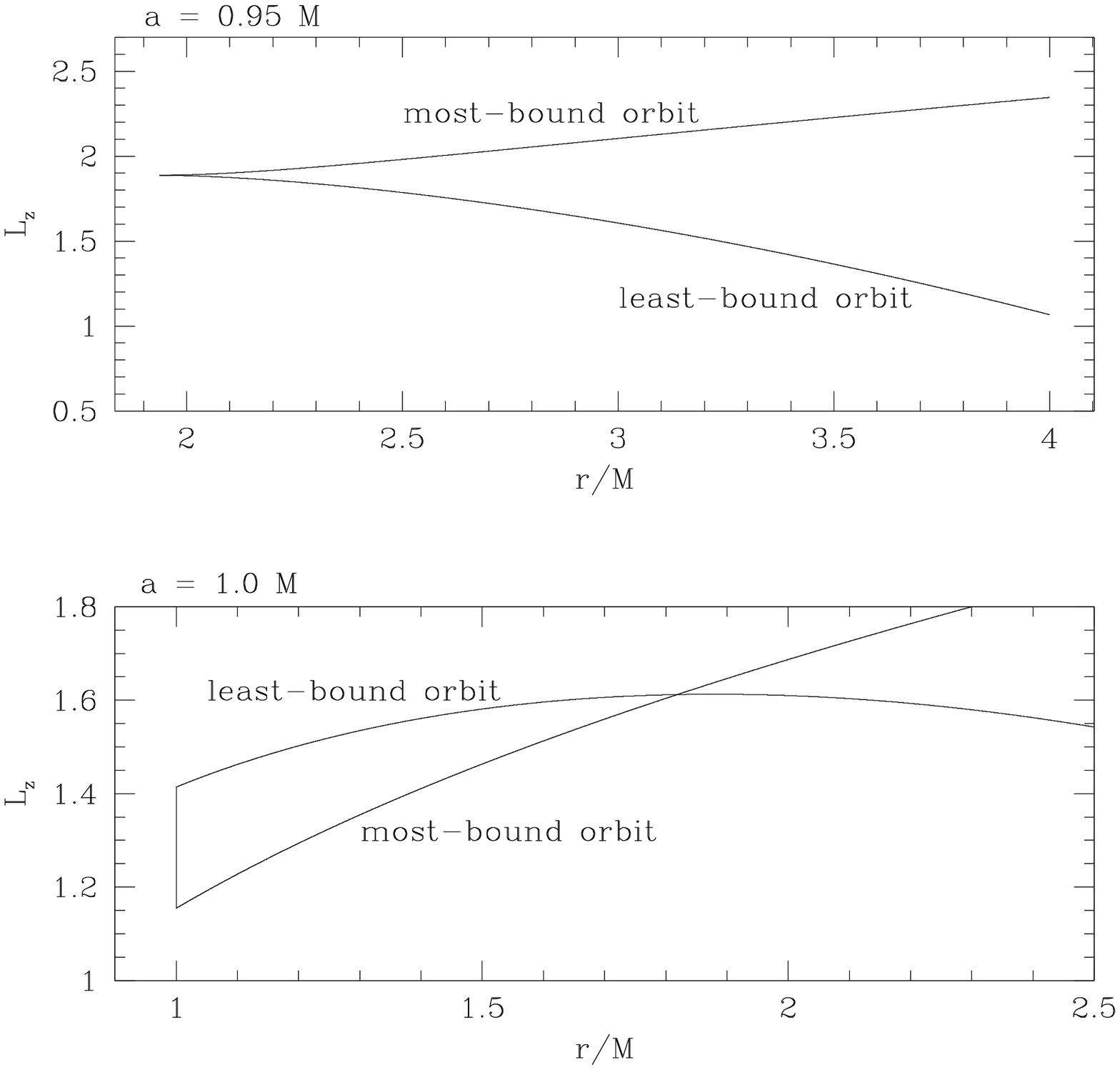, width = 16 cm}
\caption{
\label{fig:motivate_nhs}
Angular momentum $L_z$ for the most-bound and least-bound circular
orbits, as functions of radius.  The upper panel is for $a = 0.95M$,
the lower for $a = M$.}
\end{center}
\end{figure}

Figure {\ref{fig:motivate_nhs}} illustrates the NHS region, and
contrasts it with the ``usual'' behavior of Kerr orbits.  It plots
$L^{\rm mb}_z$ and $L^{\rm lb}_z$ for black holes with $a = 0.95 M$
(top panel) and $a = M$ (bottom panel).  For $a = 0.95 M$, the
least-bound and most-bound orbits coincide at the ISCO.  Moving out in
radius, the most-bound orbit's $L_z$ grows and the least-bound orbit's
$L_z$ shrinks.  This makes sense intuitively, since the inclination
angle of the least-bound orbit grows as we move away from the ISCO.
(Eventually, it tips over completely to $\iota = 180^\circ$, and
becomes the retrograde equatorial orbit.)  The lower panel of Fig.\
{\ref{fig:motivate_nhs}} shows the behavior when $a = M$.  We see the
horizon skimming orbits at $r = M$ and the NHS orbits stretching out
to $r\simeq 1.8 M$.  At that point, $L_z^{\rm lb}$ and $L_z^{\rm mb}$
cross over.  Moving further out in radius, they behave in the
``normal'' way.

There must exist some critical spin value, $0.95 M < a_{\rm NHS} < M$,
at which NHS orbits first come into existence.  The NHS orbits are
bounded by two radii, $r_{\rm ISCO}$ and $r_{\rm cross}$ [where the
angular momentum of the least-bound and most-bound orbits cross:
$L_z^{\rm mb}(r_{\rm cross}) = L_z^{\rm lb}(r_{\rm cross})$].
Decreasing $a$ must decrease the size of the NHS region --- $r_{\rm
ISCO}$ and $r_{\rm cross}$ approach one another.  The spin at which
these radii coincide and the NHS region vanishes is $a_{\rm NHS}$.

Recall from Eq.\ ({\ref{eq:EofLzr}}) that there are two solutions for
orbital energy, only one of which is usually physical.  At the
crossover point, however, {\it both} of these energies must be
physical: that with the minus sign (which is larger in magnitude)
gives $E^{\rm lb}$, and vice versa.  When $r_{\rm ISCO} = r_{\rm
cross}$ the two energies likewise coincide, since the most-bound orbit
is also the least-bound orbit at the ISCO.  Thus, $a_{\rm NHS}$ is the
spin that satisfies
\begin{equation}
E_+[r_{\rm ISCO}(a_{\rm NHS}), L_z^{\rm mb}, a_{\rm NHS}] =
E_-[r_{\rm ISCO}(a_{\rm NHS}), L_z^{\rm mb}, a_{\rm NHS}]\;
\label{eq:def_anhs}
\end{equation}
[where $E_\pm$ denotes the two roots given in Eq.\ (\ref{eq:EofLzr})].
The solution to this is $a_{\rm NHS}/M = 0.952412 \cdots\;$.

\begin{figure}[ht]
\begin{center}
\epsfig{file = 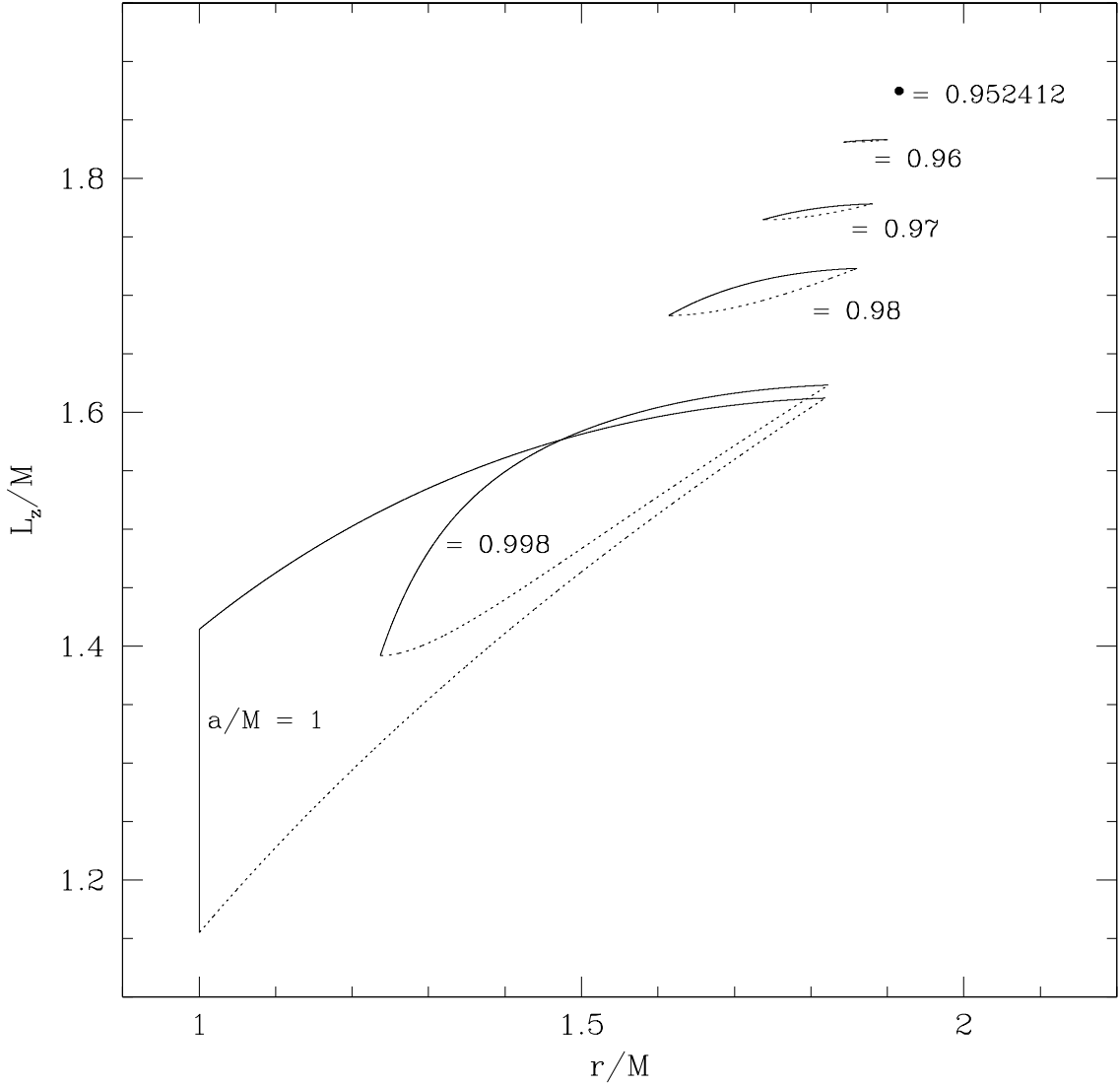, width = 16 cm}
\caption{
\label{fig:nhs_region}
The region of nearly horizon-skimming orbits for several values of the
spin.  The solid lines correspond to least-bound orbits, dotted lines
to most-bound orbits.  Notice that the region gets progressively
smaller as the spin decreases from $a = M$, vanishing altogether at $a
= a_{\rm crit} \simeq 0.952412M$.  However, the nearly
horizon-skimming region remains fairly large at least through the
vicinity of $a = 0.998 M$, which might apply to some astrophysical
black holes.}
\end{center}
\end{figure}

Figure {\ref{fig:nhs_region}} illustrates the change to the NHS region
as spin is varied, vanishing altogether when $a_{\rm NHS}$ is reached.
The size is still significant near $a = 0.998 M$.  This is
interesting, since $a = 0.998 M$ is probably the largest spin value
that we can encounter in nature {\cite{thorne_spin}}.  This opens the
possibility that NHS orbits may play a role in astrophysical
processes.

\subsection{Nearly horizon skimming orbits as seen by a ZAMO}
\label{subsec:nhs_zamo}

The NHS characteristic $\partial L_z/\partial\iota > 0$ can be better
understood by considering these orbits from the viewpoint of the zero
angular momentum observer, or ZAMO {\cite{membrane}}.  The ZAMO is the
observer that corotates with the coordinate system, such that its
angular velocity as seen at infinity is
\begin{equation}
\omega_{\rm ZAMO} = {2 a M r \over
(r^2 + a^2)^2 - a^2 \Delta\sin\theta^2}\;.
\label{eq:zamo_freq}
\end{equation}
If one imagines spacetime to be dragged into a whirlpool-like flow by
the black hole's rotation, then the ZAMO is the observer who simply
rides along with the flow.  (In accordance with this viewpoint, Wald
{\cite{wald}} calls the ZAMO the ``locally non-rotating observer''.)

The ZAMO examines orbits in its local neighborhood.  This allows it to
interpret the motion of a body in a strong-field orbit with special
relativistic formulas.  For instance, the ZAMO measures the small body
to have velocity $\vec v$ and energy $E_{\rm local} = (1 - \vec
v\cdot\vec v)^{-1/2} \equiv\gamma$.  This local energy is {\it not} a
constant of the body's motion.  It is related to the conserved energy
by the formula {\cite{membrane}}
\begin{equation}
E = \alpha E_{\rm local} + \omega_{\rm ZAMO} L_z\;,
\label{eq:Elocal_to_E}
\end{equation}
where
\begin{equation}
\alpha = {\sqrt{\Sigma\Delta\over{(r^2 + a^2)^2 -
a^2\Delta\sin\theta^2}}}\;
\label{eq:lapse}
\end{equation}
is the lapse function.  Knowing the energy and the angular momentum of
the small body then tells us the body's speed $v = \sqrt{\vec
v\cdot\vec v}$ as seen by the ZAMO.

Applying Eq.\ (\ref{eq:Elocal_to_E}) shows that a body's speed is
smallest in the most-bound orbit and highest in the least-bound orbit,
varying smoothly between the two.  This is due to the dragging of
spacetime by the hole's spin.  Let us contrast the prograde and
retrograde orbits, $\iota = 0^\circ$ and $\iota = 180^\circ$.  The
prograde orbit is moving ``downstream'': part of the motion needed to
keep it in orbit is provided by the dragging of inertial frames.  It
can orbit with relatively small velocity.  The retrograde orbit, by
contrast, must ``swim upstream'': it must overcome the dragging of
inertial frames on top of the motion needed to stay in orbit.  It
therefore is more energetic than the prograde orbit.  The angle
$\iota$ smoothly varies the orbit between these extremes, so that
larger $\iota$ corresponds to larger energy (and larger speed).

The velocity of a body in a non-equatorial orbit has components in
both the $\theta$ and $\phi$ directions.  The $\theta$ component goes
to zero, however, at the orbit's turning points $(\theta_{\rm
max},\theta_{\rm min})$, when it reverses in $\theta$.  (Formulas for
computing $\theta_{\rm max/min}$ can be found in Ref.\
{\cite{paperI}}; they are just the angles at which $d\theta/d\tau = 0$
[cf.\ Eq.\ (\ref{eq:thetadot})].)  At these two points, the velocity
is purely along $\phi$, and the body's motion is fully described by
the component $v_{\hat\phi} = \vec v\cdot {\vec e}_{\hat\phi}$ (where
${\vec e}_{\hat\phi}$ is the $\phi$-component of the orthonormal basis
that the ZAMO uses to make measurements).  Evaluating Eq.\
(\ref{eq:Elocal_to_E}) at $\theta_{\rm max/min}$ and writing $E_{\rm
local} = \gamma$ gives a condition for $v_{\hat\phi}$ at the turning
points.

The top panel of Fig.\ {\ref{fig:vphi_pphi}} shows
$v_{\hat\phi}(\theta_{\rm max/min})$ for the most-bound and
least-bound orbits at several interesting spins.  As expected, bodies
move quite a bit faster in the least-bound orbit than in the
most-bound orbit.  Perhaps more interestingly, $v_{\hat\phi}$ becomes
substantially smaller towards the innermost orbits.  This is because
of the ``freezing'' of physics near the hole's event horizon; see
Ref.\ {\cite{membrane}} (particularly Secs.\ IIC1 and IIIA4) for
further discussion.  Close to the horizon, a body's motion locks to
the hole's spin, and it is dragged into rigid corotation.  This
horizon locking causes $v_{\hat\phi}(r)$ to peak at $r \sim 1.5 M$:
the ``Keplerian'' tendency of a body to move faster as it moves inward
dominates at large radii (asymptoting to $\sqrt{M/r}$ in the weak
field), but is overwhelmed as the body locks onto the horizon close to
the hole.

Consider next the orbit's angular momentum $L_z$.  The ZAMO sees $L_z$
as the product of a radius of gyration $\varpi$ and a locally measured
azimuthal momentum $p_{\hat\phi}$ {\cite{membrane}}:
\begin{equation}
L_z = \varpi p_{\hat\phi} = \varpi \gamma v_{\hat\phi}\;.
\label{eq:angmom_zamo}
\end{equation}
The $\gamma$ factor causes $p_{\hat\phi}$ to be even more strongly
peaked than $v_{\hat\phi}$; see the lower panel of Fig.\
{\ref{fig:vphi_pphi}}.

\begin{figure}[ht]
\begin{center}
\epsfig{file = 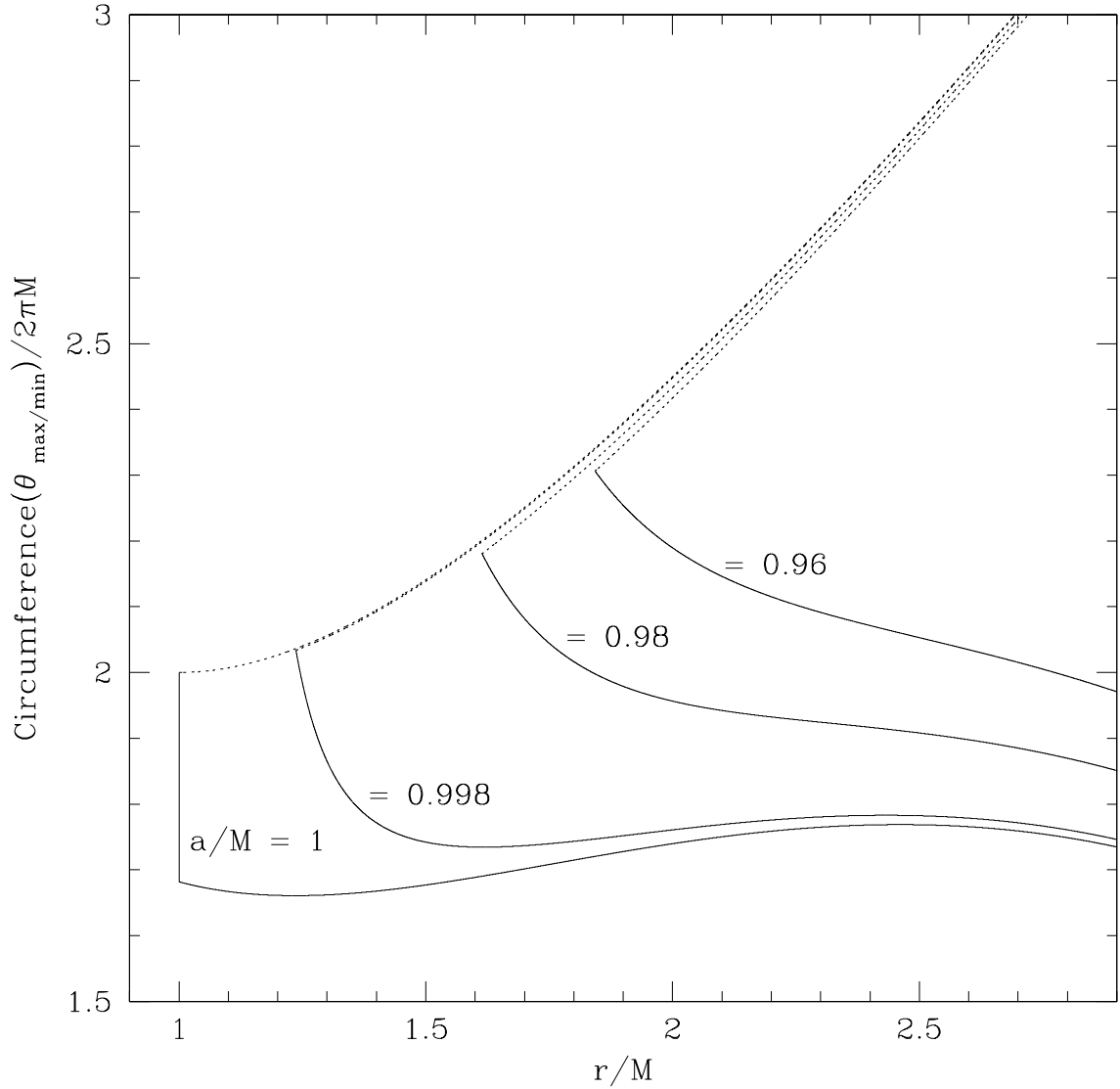, width = 16 cm}
\caption{
\label{fig:radgyration}
The radius of gyration $\varpi = \mbox{Circumference}(\theta_{\rm
max/min})/2\pi$ for the most-bound orbit (dotted curve) and
least-bound orbit (solid curve).  Notice that $\varpi$ barely varies
for the least-bound orbit over this range.  This is essentially
because the change in $\theta_{\rm max/min}$ compensates for the
change in radius such that the circumference of the least-bound orbit
remains roughly constant.}
\end{center}
\end{figure}

The radius of gyration is a purely geometric quantity.  It is just the
circumference of the ZAMO's constant $r$, constant $\theta$ orbit,
divided by $2\pi$:
\begin{eqnarray}
\varpi &=& {\cal C}(\theta_{\rm max/min})/2 \pi
= {1\over2\pi}\int_0^{2\pi} \sqrt{g_{\phi\phi}(r, \theta_{\rm
max/min})}\,d\phi
\nonumber\\
&=& \sqrt{{(r^2 + a^2)^2 - a^2 \Delta \sin\theta_{\rm
max/min}^2}\over{r^2 + a^2 \cos\theta_{\rm
max/min}^2}}\,\sin\theta_{\rm max/min}\;.
\end{eqnarray}
This function is plotted in Fig.\ {\ref{fig:radgyration}}.  There are
no surprises here.  In the most-bound orbit, $\varpi$ increases
monotically with orbital radius, asymptoting to $r$ at large radius.
The least-bound orbit is more interesting: $\varpi$ is smaller (not
surprising, since it is a tilted orbit) and is nearly flat as a
function of radius, at least over the range of NHS orbits.  This near
flatness is due to the orbit's increasing tilt: $\theta_{\rm max/min}$
of the least-bound orbit changes such that the circumference of the
ZAMO's orbit at $\theta_{\rm max/min}$ remains nearly constant.

Since $L_z$ is just the product of the curves shown in Figs.\
{\ref{fig:vphi_pphi}} and {\ref{fig:radgyration}}, any unusual
features in the behavior of $L_z$ must arise from features in
$p_{\hat\phi}$ and $\varpi$.  Considering these two figures, we
immediately see why NHS orbits have $L_z^{\rm lb} > L_z^{\rm mb}$: the
least-bound orbit has so much more linear momentum than the most-bound
orbit that it compensates for its smaller radius of gyration.  The
linear momentum is so much larger, in turn, because the most-bound
orbit is strongly locked to the spin of the black hole.  The
least-bound orbit is also locked for $r$ very close to $r_{\rm ISCO}$.
However, it does not lock as strongly: being so energetic, the
least-bound orbit only locks as the very innermost orbits are
approached.  Hence, NHS orbits exist because very close orbits are
forced to move in rigid corotation with the event horizon.

\section{Application: evolution under gravitational-wave emission}
\label{sec:radiate}

As discussed in the Introduction, binary systems consisting of a small
body spiraling into a massive black hole are one of the more
anticipated sources of gravitational waves for space-based detectors
such as LISA.  The NHS region is still rather large for $a \simeq
0.998M$ (cf.\ Fig.\ {\ref{fig:nhs_region}}), indicating that there
might be plenty of time for the properties of NHS orbits to influence
the gravitational-wave signal of these sources.

In the extreme mass ratio, radiation reaction should operate
adiabatically.  In other words, the timescale for gravitational-wave
emission to change an orbiting body's parameters ($r$, $E$, $L_z$,
$Q$) should be significantly longer than an orbital period: $T_{\rm
orb}/\tau_{\rm GW} \ll 1$.  (More careful discussion of this point can
be found in Ref.\ {\cite{paperI}}.)  Because the change in these
constants is very slow, the body's motion is well-approximated as
geodesic over small time intervals.  It is thus useful to regard the
body's true, radiatively evolving trajectory as motion through a
sequence of geodesic orbits.  For circular orbits, we regard the
body's inspiral trajectory as the evolution of its radius and
inclination angle.

A simple argument shows that a body spiraling through the NHS region
should behave rather differently from a body spiraling through
``normal'' orbits.  At a given moment, a circular orbit can be
represented as a point on the $(r,L_z)$ plane.  For example, a point
on either the top or the bottom panels of Fig.\
{\ref{fig:motivate_nhs}} lying between the most-bound and least-bound
orbit curves represents a physically allowed orbit.  Radiation
reaction will tend to drive this point downward and to the left in
this figure: gravitational-wave emission shrinks an object's radius
and reduces $L_z$.  For a ``normal'' orbit, as in the top panel of
Fig.\ {\ref{fig:motivate_nhs}}, the body is pushed toward less-bound
orbits, increasing $\iota$.  This behavior was predicted by weak-field
calculations {\cite{fintan_circ,fintan_gen}} and was recently
confirmed in the strong field {\cite{paperI}}.  The evolution is in
the opposite direction inside the NHS region --- radiation emission
pushes the body toward the sequence of most-bound orbits, so that
$\iota$ decreases.

Decreasing inclination angle could have important consequences for the
gravitational-wave signal.  Inclined circular orbits emit waves
characterized by harmonics of $\Omega_\phi = 2\pi/T_\phi$ and
$\Omega_\theta = 2\pi/T_\theta$ (where $T_\phi$ is the period of a
full range of $\phi$ motion, and $T_\theta$ is the period of $\theta$
motion) {\cite{paperI}}.  Roughly speaking, the $\theta$ motion
modulates the gravitational waveform.  The waves emitted by a body in
an equatorial orbit, on the other hand, depend only on $\Omega_\phi$
--- there is no motion in $\theta$, so the amplitudes of
$\Omega_\theta$ harmonics vanish.  As the orbiting body moves toward
the equatorial plane, the importance of the $\Omega_\theta$ harmonics
decreases, simplifying the waveform.  Observing such simplification
would be a clear signal that the small body is evolving toward the
equatorial plane.

A rigorous analysis of NHS evolution requires a radiation reaction
formalism good deep in the strong field.  Such a formalism,
specialized to the adiabatic evolution of circular orbits, is given in
Ref.\ {\cite{paperI}}.  The code described there solves the Teukolsky
equation {\cite{teukolsky}} to calculate the flux of $E$ and $L_z$ to
infinity and down the event horizon.  It then applies the proof that
circular orbits remain circular (changing only their radius and
inclination angle) when they evolve adiabatically under radiation
emission {\cite{mino_circ,fintan_gen,dan_amos}} to compute the change
in $Q$.  From this information, it is simple to obtain the change in
$r$ and $\iota$.  By computing $dr/dt$ and $d\iota/dt$ at a large
number of points, it is not difficult to compute the inspiral
trajectory followed by a body that starts in an inclined, circular
orbit.  An algorithm and code for computing such trajectories will be
presented at a later date {\cite{paperII}}.

\begin{figure}[ht]
\begin{center}
\epsfig{file = 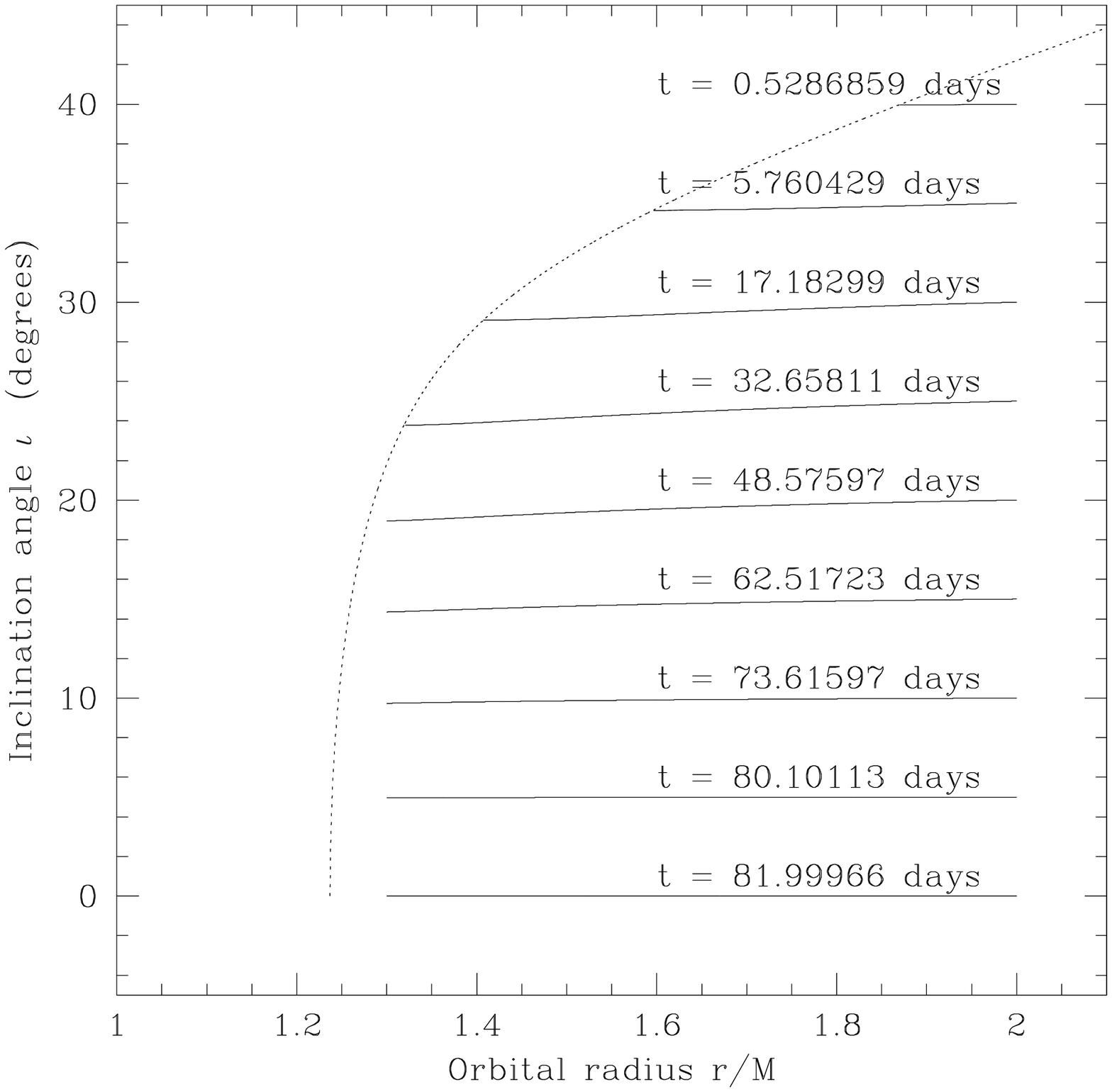, width = 16 cm}
\caption{
\label{fig:nhs_seq}
The evolution of radius and inclination angle for a small body
inspiraling through the NHS region.  The small body has $\mu =
1\,M_\odot$, the large black hole has $M = 10^6\,M_\odot$ and spin $a
= 0.998 M$.  The inclination angle decreases in this region, opposite
to the usual behavior.  The amount of decrease is, however, rather
small.  [The gap at the end of the inspiral trajectories for $\iota\le
20^\circ$ is because of computational limitations: it is
computationally expensive to generate a very dense mesh of
$(dr/dt,d\iota/dt)$ data close to the ISCO.]}
\end{center}
\end{figure}

Figure {\ref{fig:nhs_seq}} shows some trajectories in the NHS region
for the inspiral of a $1\,M_\odot$ compact body into a $10^6\,M_\odot$
Kerr black hole.  The hole has spin $a = 0.998 M$.  The inspiral time
for most trajectories is significant.  The number of orbits around the
spin axis of the black hole by the body varies from $N_\phi\simeq
73000$ for the inspiral beginning at $\iota = 0^\circ$ to
$N_\phi\simeq 400$ for the inspiral beginning at $\iota = 40^\circ$.
The number of gravitational-wave cycles is roughly proportional to
$N_\phi$, indicating that waves from the NHS region could contribute
substantially to a measured signal.  The inspiral duration scales with
$M^2/\mu$, and the number of orbits with $M/\mu$.  These quantities
should be of interesting magnitude for a reasonably wide range of
masses, at least for spin $a = 0.998 M$.  (As the spin of the black
hole is dialed down to $a = a_{\rm NHS}$, the NHS orbits lose
significance and inspiral through them becomes irrelevant.)
Gravitational waves produced in the NHS region will form a significant
part of the data measured from inspirals into rapidly rotating black
holes.

Figure {\ref{fig:nhs_seq}} also indicates that the signature of NHS
inspiral is not very large.  Although $\iota$ does decrease, as
expected, the degree of decrease is quite small.  For the inspirals
shown, the largest change in $\iota$ is for the trajectory that starts
at $\iota = 25^\circ$: the final inclination is $\iota = 23.8^\circ$.
The change $\delta\iota = 1.2^\circ$ is a rather paltry effect, not at
all a robust signature of the strong field.

It seems that the unique signature of orbital evolution through the
NHS region is not likely to be of observational interest.  Before
dismissing the idea entirely, let us speculate for a moment on how
eccentricity might affect this conclusion.  Using a weak-field
radiation reaction force, Ryan has calculated the leading order
evolution of the parameters of an inclined, eccentric orbit of a Kerr
black hole due to gravitational-wave emission {\cite{fintan_gen}}.
His results indicate that the eccentricity $e$ can significantly
magnify the rate at which $\iota$ changes:
\begin{equation}
\left.{d\iota\over dt}\right|_{e \ne 0} = \left(1 - e^2\right)^{-7/2}
\left(1 + {73\over 24}e^2 + {37\over 96}e^4\right)
\left.{d\iota\over dt}\right|_{e = 0}\;.
\label{eq:eccentric_effect}
\end{equation}
This relation is derived under the assumption that $r \gg M$, so there
is no reason to believe it holds in the NHS region.  However, it could
be indicative of the effect that eccentricity has in the strong field.
Notice that as eccentricity varies over the range $0 \le e \le 0.8$ (a
reasonable range for sources that LISA is likely to measure
{\cite{eccentricity_comment}}), the factor magnifying $d\iota/dt$
increases from $1$ to about $100$.  One can speculate that $d\iota/dt$
will be magnified by a similar factor (at least to order of magnitude)
under strong-field radiation reaction.  At this time, there is no way
to tell.  Hopefully, strong-field radiation reaction programs
{\cite{rad_react_cabal,wolfgang_eanna}} will be able to model the
evolution of generic Kerr orbits soon.  If this speculation proves
correct, then the gravitational-waves emitted by inspiral through the
NHS region will contain a very strong signature of the Kerr metric
strong field.

\section{Conclusion}
\label{sec:conclusion}

The result of this analysis is somewhat equivocal.  I have shown that
the ``horizon skimming'' orbits found by Wilkins are a subgroup of a
larger family of orbits, the ``nearly horizon skimming'' orbits
existing around any black hole with spin $a \agt 0.952412 M$.  The
signature characteristic of these NHS orbits is that, at fixed orbital
radius, the $z$-component of orbital angular momentum increases as the
orbit's tilt increases, in opposition to the ``normal'' behavior.
This has the consequence that radiation emission, which carries
angular momentum away from the orbit, tends to drive the system into
an equatorial orbit --- the system's inclination angle {\it decreases}
rather than increases.

\begin{figure}[ht]
\begin{center}
\epsfig{file = 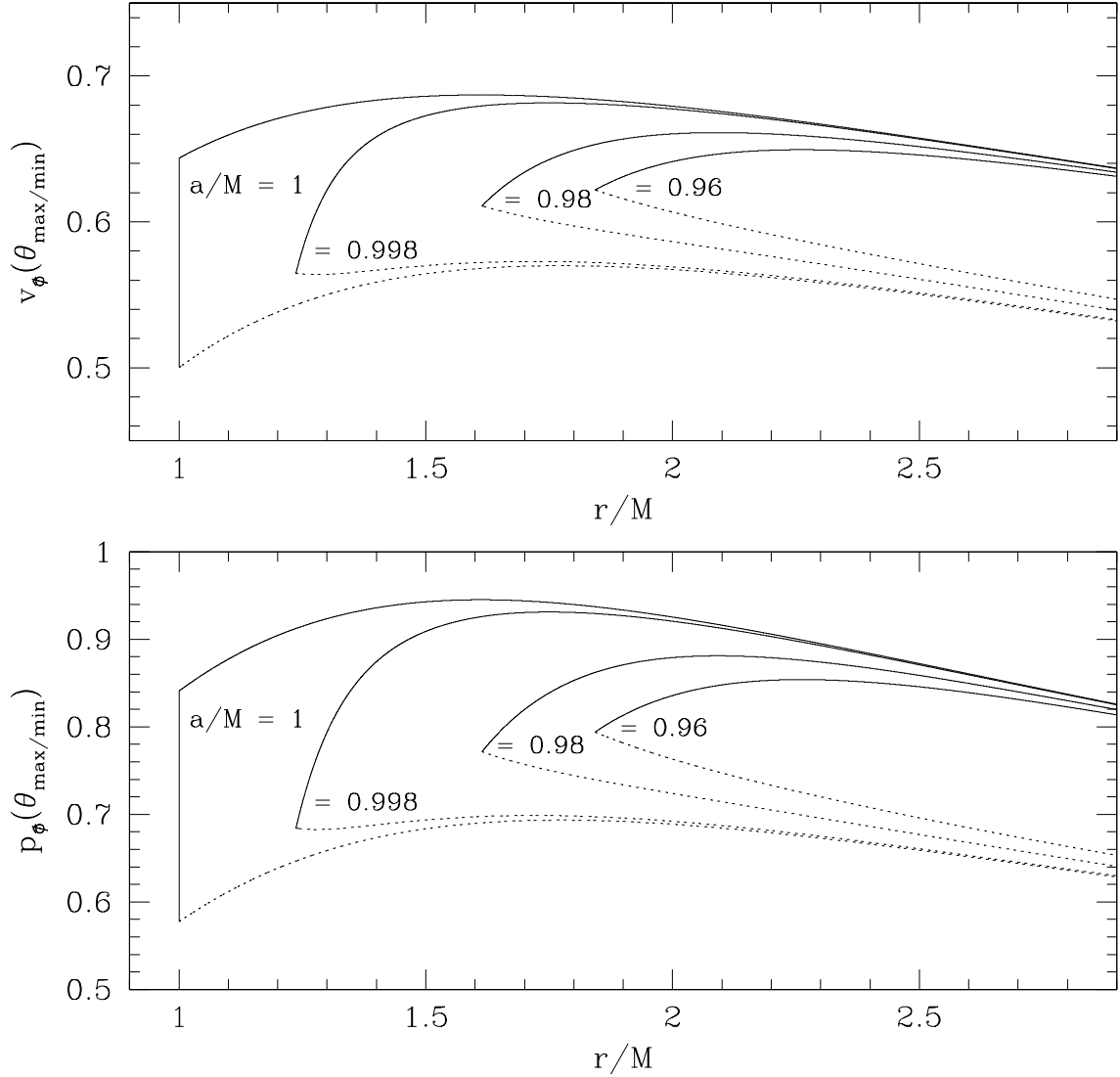, width = 16 cm}
\caption{
\label{fig:vphi_pphi}
The azimuthal velocity and momentum at the $\theta$ turning points of
an orbiting compact body, as seen by a ZAMO.  The top panel displays
$v_{\hat\phi}$ for the most-bound orbit (dotted curve) and least-bound
orbit (solid curve) at several spin values; the bottom panel likewise
displays $p_{\hat\phi}$.  As expected, a body in the least-bound orbit
moves substantially faster than a body in the most-bound orbit.
Notice that the motion becomes slower as the ISCO (the innermost
orbit) is approached.  This is because the ISCO is close to the
horizon at these spin values.  Orbits that come close to the horizon
become locked to the rotation of the black hole.  This close to the
horizon, locking is substantial.  This locking is reponsible for the
peaks in these functions: moving inward along a sequence of orbits, a
body first orbits more quickly, but then slows as its motion locks to
the spin.  The momentum peak is magnified (note the different vertical
scales in the two panels) because $p_{\hat\phi} = \gamma v_{\hat\phi}
= v_{\hat\phi}(1 - v_{\hat\phi}^2)^{-1/2}$.}
\end{center}
\end{figure}

The fact that radiation emission drives these orbits toward the
equator is a consequence of the motion of bodies in the strong field
of Kerr black holes.  As a body spirals in from large radius, its
orbital speed initially increases, in accordance with Keplerian
intuition that a body in a circular orbit of radius $r$ has a speed $v
= \sqrt{M/r}$.  As the body gets very close to the black hole,
however, its dynamics become dominated by the nearness of the hole's
event horizon.  All physical processes become ``locked'' to the hole
as the horizon is approached.  Eventually the motion of orbiting
bodies synchronizes with the hole's spin.  The least-bound orbit does
not lock as quickly as the the most-bound orbit: close to the horizon,
a less-bound orbit moves quite a bit faster than a highly bound orbit,
not locking to the hole's spin until the very innermost orbits are
reached.  This effect is shown in Fig.\ {\ref{fig:vphi_pphi}}.

As an inspiraling body's inclination decreases, the modulation of its
gravitational waveform by the $\theta$ motion decreases.  Measuring
this decreasing modulation could be an observational hallmark of the
Kerr strong field.  Unfortunately, at the moment the most mature
formalism for computing the effects of gravitational radiation
reaction on the orbits of small bodies can only tackle the evolution
of circular, inclined orbits.  In this restricted case, the total
change in inclination angle as a body evolves through the NHS region
is quite small --- at most, the orbital inclination changes by about
$1 - 2$ degrees.  Although I have not examined the effect of this
change on the waveform, it is hard to imagine it will be particularly
marked.

It is extremely unlikely that realistic extreme mass ratio inspirals
will be circular.  When such binaries initially form, their
eccentricities are likely to be rather close to $1$
{\cite{sigurdsson_rees,sigurdsson}}, and remain significant by the
time that the binary enters the sensitivity band of LISA.  Weak field
calculations suggest that the rate by which the inclination angle
changes might be magnified by a rather large factor.  If we take the
leading order results at face value [cf.\ Eq.\
(\ref{eq:eccentric_effect})], the change might be so strong that the
orbit is driven nearly into the equatorial plane.

Because the NHS region is so deep in the strong field, however, it is
a mistake to infer too much from leading order effects.  Further
progress will require a scheme to calculate radiation reaction in the
strong field for {\it generic} Kerr orbits --- orbits that are
inclined and eccentric.  This may require local radiation reaction
forces {\cite{rad_react_cabal}}, or a clever scheme for extracting the
change in the Carter constant from the radiation flux
{\cite{wolfgang_eanna}}.  The fact that observations might be able to
detect a strong signature of the Kerr strong field will hopefully
motivate future progress.

\acknowledgements

This work grew out of a more general program to compute waveforms and
radiation reaction sequences from extreme mass ratio circular orbit
inspirals.  That work, in turn, has benefitted greatly from
conversations with Lior Burko, Curt Cutler, Dan Kennefick, Sam Finn,
Lee Lindblom, Sterl Phinney, and Kip Thorne.  The package {\sc
Mathematica} was used to aid some of the calculations; all plots were
produced using the package {\sc sm}.  This research was supported by
NSF Grant AST-9731698 and NASA Grants NAG5-7034 and NAGW-4268 at
Caltech, and NSF Grant PHY-9907949 at the ITP.

\eject

\begin{references}
\bibitem{eccentricity_comment} This can be seen by taking a binary
	with initial parameters as described in Refs.\
	{\cite{sigurdsson_rees,sigurdsson}} and using, for instance,
	formulas in {\cite{peters_mathews}} to estimate the
	eccentricity when the system enters the LISA frequency band.
\end{references}
\end{document}